\documentclass{PoS}

\title{Anomaly cancellation for anisotropic lattice fields with extra dimensions}

\ShortTitle{Anomaly cancellation on the lattice with extra dimensions}

\author{\speaker{Stam Nicolis}\\
        CNRS--Laboratoire de Math\'ematiques et Physique Th\'eorique (UMR 6083)\\
	F\'ed\'eration de Recherche ``Denis Poisson'' (FR 2964)\\
	Universit\'e ``Fran\c{c}ois Rabelais'' de Tours\\
	Parc Grandmont, 37200 Tours, France\\
        E-mail: \email{Stam.Nicolis@lmpt.univ-tours.fr}}

\abstract{The current flow from the bulk is due to the anomaly 
on the brane--but the absence of current flow is not, necessarily,
  due to anomaly cancellation, but to the absence of the chiral zero modes 
themselves, due to the existence of the layered phase. This can be understood
in terms of the difference between the Chern--Simons terms in three and five
dimensions. Thus the anomaly cancellation in four dimensions, which is
essential for shielding the boundary from quantum effects within the bulk, 
 makes sense only along the transition line between the layered and the 
Coulomb phase, which, in turn, requires the presence of a compact $U(1)$
factor for the gauge group.  }

\FullConference{The XXVII International Symposium on Lattice Field Theory - LAT2009\\
		 July 26-31 2009\\
		 Peking University, Beijing, China}

\begin{document}

Domain wall~\cite{kaplan} and overlap~\cite{neuberger} fermions allow the
realization of chiral symmetry on the lattice~\cite{luescher}. The key ingredients, that lead
to evading the no--go theorem~\cite{nielsen_ninomiya}, are the defects
that extend in extra dimensions. The chiral zero modes are localized along
appropriate boundaries. For a consistent quantum theory we would like that the
boundary theory be protected from quantum effects within the bulk. 

Since the fermions are  coupled to gauge fields, there will
generically 
be anomalies~\cite{callan_harvey}. These appear as current flow to and from
the bulk. They also appear as ``edge currents''. Therefore anomalies should be
properly cancelled~\cite{bim}. Anomaly cancellation has been studied for the
chiral Schwinger model~\cite{jansen}, i.e.  a three-dimensional bulk with 
two-dimensional boundaries, but
not for the five-dimensional case. The object of this note is to point out
some essential differences between the two cases. These differences are also 
relevant for attempts at detecting extra dimensions~\cite{shaposhnikov}. 

\begin{figure}[thp]
\begin{center}
\includegraphics[scale=0.4]{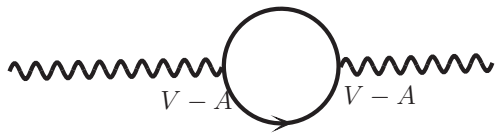}
\end{center}
\caption[]{The contribution to the anomaly in $D=2$.}
\label{D2An}
\end{figure}

\begin{figure}[thp]
\begin{center}
\includegraphics[scale=0.4]{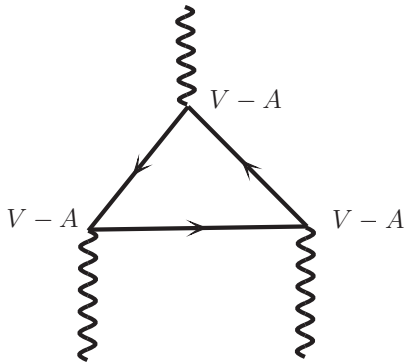}
\end{center}
\caption[]{The contribution to the anomaly in $D=4$.}
\label{D4An}
\end{figure}

In three dimensions the relevant contribution to the anomaly is the vacuum
polarization diagram (cf. Fig.~\ref{D2An})
which is well-known to provide a ``topological'' mass term for the gauge
field~\cite{DJT,coleman_hill}. In five dimensions the relevant contribution is
the well-known triangle diagram (cf. Fig.~\ref{D4An})
which does {\em not} provide a mass for the gauge field--it is an interaction
term. In fact this is the crucial difference: in three dimensions the anomaly
is a ``soft'' term, whereas in five dimensions it is a ``hard'' term. This was
in fact noticed in refs.~\cite{adler,preparata_weisberger} for the
four--dimensional anomaly in that the renormalization constant for the axial
current could not be expressed in terms of the renormalization constant of
mass term and that of the wave function renormalization constant of the
fermion, but was an independent quantity. 

\begin{figure}[thp]
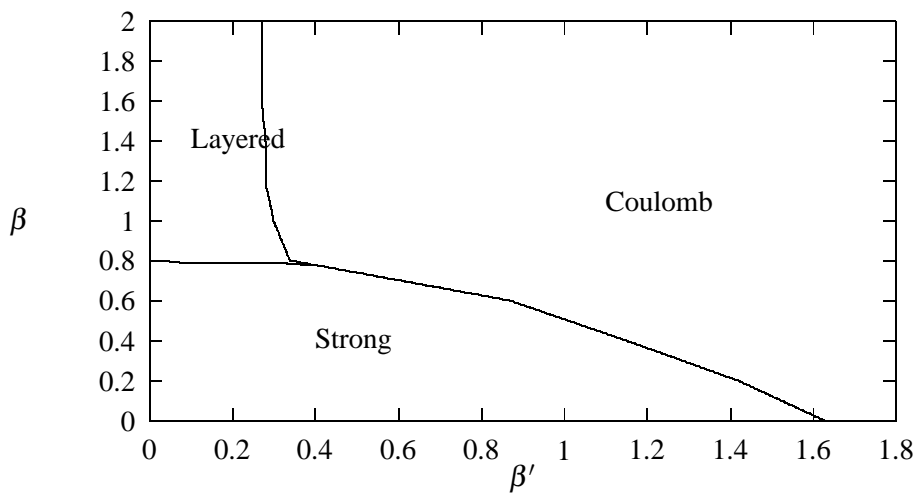

\input phadi5.tex
\caption[]{The phase diagram of the five--dimensional $U(1)$ theory in the 
$\beta-\beta'$ plane.}
\label{phadi5}
\end{figure}

In fact it is possible to understand this in another way. The anisotropy of
the defects implies~\cite{kaplan} that the gauge couplings will generically be
different along the boundary ($\beta$) and towards the bulk ($\beta'$). The
action on the lattice will then take the form (for the five--dimensional case)
\begin{equation}
\label{latt_action}
S = \beta\sum_n\sum_{\mu<\nu}(1-\mathrm{Re} U_{\mu\nu}(n)) + 
    \beta'\sum_n\sum_\mu(1-\mathrm{Re} U_{\mu5}(n)) + S_\mathrm{fermions} 
\end{equation}
In the three--dimensional case the isotropic theory, $\beta=\beta'$ is
renormalizable (in fact it is super-renormalizable). This means that, once
anomalies are cancelled, the two--dimensional theory is protected from quantum
bulk effects. In the five--dimensional case, however, the isotropic theory is
{\em not} renormalizable-it cannot be made insensitive to cutoff effects,
which renders the decoupling of its boundaries questionable. 
However, the anisotropic theory may have a
layered phase~\cite{fu_nielsen}:  and along the transition line between this
phase and the bulk Coulomb phase it seems that a continuum theory could be
defined, since this transition seems to
 be second order~\cite{HKAN,farakos_vrentzos} (cf. Fig.~\ref{phadi5})
Therefore it is for gauge couplings along this line that the anomaly
cancellation can  be realized. Indeed, the current, that carries the anomaly
flow from and to the bulk plays a double r\^ole: on the one hand it is a
component of the bulk vector current; on the other hand it is the axial
current density from the point of view of the boundary. The chiral charge on
the boundary has absolute value that of the (vector) bulk charge and sign the
chirality at the boundary. Thus these models have a {\em gauged} chiral
symmetry. 

\begin{figure}[thp]
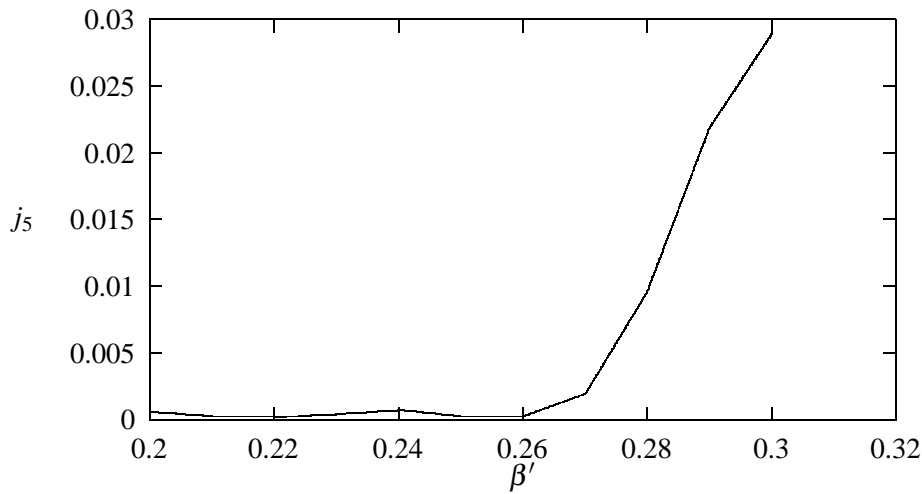

\input J5.tex
\caption[]{The current along the extra dimension, for $\beta=1.2$, as a
  function of $\beta'$.}
\label{current}
\end{figure}

This current can indeed be computed~\cite{HKAN} and displays the following
behavior (cf. Fig.~\ref{current})
It is noteworthy that this current vanishes within the layered phase and is
non-zero in the bulk Coulomb phase. The reason it does not vanish in the
Coulomb phase is a manifestation of the anomaly--the reason it vanishes in the
layered phase is not anomaly cancellation, but, rather, that there isn't any
chiral zero mode that can absorb or emit the charge~\cite{KANP}. 
 
The model we studied was based on the compaact formulation of the 
$U(1)$ gauge group. The reason is that this group is the only one that
presents both a Coulomb phase and a confining phase, that are separated by a
phase transition. This is essential to the possibility of defining the layered
phase intrinsically~\cite{nicolis07}. Anisotropic Yang--Mills theories are
harder to study (cf. ref.~\cite{kurkela} for a study of the decoupling effects
in a model with $SU(2)$ gauge group; it will
be interesting to compare this with the case of 
$U(2)=SU(2)\otimes U(1)$, which should
provide a much cleaner signal, since the $U(1)$ factor will trigger the
layered to Coulomb phase transition along which the anomaly cancellation
should be realized. Studies with $SU(3)$ have been attempted~\cite{fu} but did
not take into account this difference, so it is hard to see exactly how the
layer is defined.). 

In conclusion, the difference between the anomaly terms in three and five
dimensions is crucial in understanding the difference in the physics between
the two cases.  In three dimensions the anomaly term is a mass term and
 both the bulk and the boundary theories are renormalizable at the 
isotropic point, where anomaly cancellation can be imposed. 
 In five dimensions the anomaly term is an interaction vertex 
(whose, dimensionless,  coefficient may be related to the ratio of the, 
dimensionful, coupling constants in the boundary and towards the bulk). 
Anomaly cancellation cannot take place along the isotropic line,since the 
bulk theory is not renormalizable there--the layered phase, 
however, leads to the possibility of shielding the boundary from the bulk 
along a line of strongly coupled fixed points, that may be also
relevant in other contexts (e.g.~\cite{pallante}).

\acknowledgments
It is a pleasure to acknowledge discussions with S. M. Catterall, Ph. de
Forcrand, L. Giusti and M. Panero.

\end{document}